\documentclass[]{article}

\usepackage{amsmath}
\usepackage{amssymb}
\usepackage{tikz}
\usepackage{graphicx}
\usepackage[footnotesize,hang]{caption}
\usepackage{subfig}
\usepackage{color}
\usepackage{rotating}
\usepackage{mathtools}
\usepackage{xcolor}

\usepackage[margin=1in]{geometry}

\usetikzlibrary{decorations.pathmorphing}

\newcommand{\pdiff}[2]{\frac{\partial #1}{\partial #2}}

\newcommand{\eps}{\epsilon}

\renewcommand{\i}{\mathrm{i}}
\newcommand{\e}{\mathrm{e}}

\def\XXint#1#2#3{{\setbox0=\hbox{$#1{#2#3}{\int}$}
\vcenter{\hbox{$#2#3$}}\kern-.5\wd0}}

\title{Generalized Solitary Waves in a Finite-Difference Korteweg-de Vries Equation}
\author{N. Joshi$^1$\footnote{Electronic address: nalini.joshi@sydney.edu.au;} and C. J. Lustri$^2$\footnote{Electronic address: christopher.lustri@mq.edu.au; corresponding author}}
\date{%
    $^1$School of Mathematics and Statistics, F07, The University of Sydney, New South Wales 2006, Australia\\%
    $^2$Department of Mathematics and Statistics, 12 Wally's Walk, Macquarie University, New South Wales 2109, Australia\\[2ex]%
}                                     

\begin{document}
\maketitle

\abstract{Generalized solitary waves with exponentially small non-decaying far field oscillations have been studied in a range of singularly-perturbed differential equations, including higher-order Korteweg-de Vries (KdV) equations. Many of these studies used exponential asymptotics to compute the behaviour of the oscillations, revealing that they appear in the solution as special curves known as Stokes lines are crossed. Recent studies have identified similar behaviour in solutions to difference equations. Motivated by these studies, the seventh-order KdV and a hierarchy of higher-order KdV equations are investigated, identifying conditions which produce generalized solitary wave solutions. These results form a foundation for the study of infinite-order differential equations, which are used as a model for studying lattice equations. Finally, a lattice KdV equation is generated using finite-difference discretization, in which a lattice generalized solitary wave solution is found.}

\section{Introduction}

Following widespread interest in in the existence of generalized solitary waves (GSW) in Korteweg-de Vries (KdV) type continuous models, we show for the first time that GSW appear in a discrete KdV equation arising from finite difference discretization. This is related to an infinite-order partial differential equation, and so we investigate related phenomena in fifth- and seventh-order KdV equations to lay the groundwork for our study. Following this line of reasoning, we then consider a finite difference discretization of the unperturbed KdV equation, producing a lattice equation that is given in \eqref{1.fdkdv}. While this system does not immediately appear to have an obvious singular perturbation, singularly perturbed terms are in fact introduced into the system by the discretization process itself. Adapting the discrete exponential asymptotic methods used in \cite{Joshi4,Joshi5, King4}, we provide the first example of GSW in a lattice.

\subsection{Background}

\begin{figure}
\centering
\includegraphics{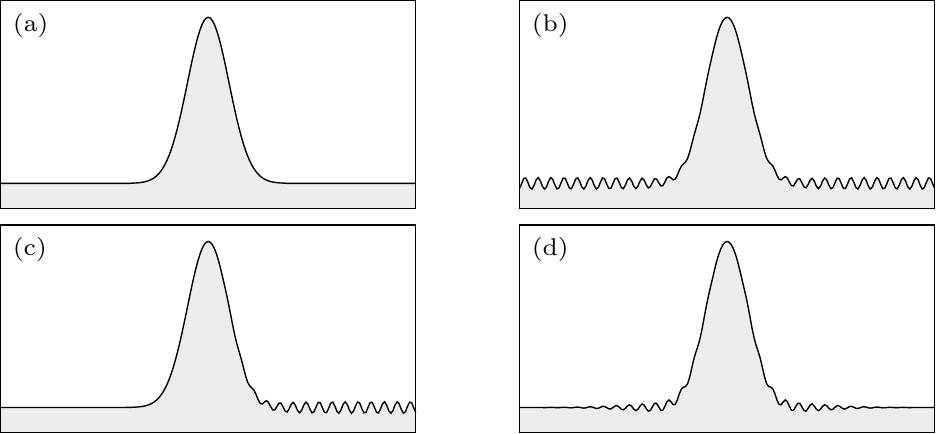}
%
%
%
%
%
%
%

\caption{Schematic of solitary and generalized solitary waves. Figure (a) depicts a spatially-localized solitary wave. Figures (b) and (c) depict a two- and one-sided generalized solitary wave respectively. Figure (d) depicts a solitary wave with exponentially small oscillations that decay away from the wave core.}\label{F:GSW}
\end{figure}

Generalized solitary waves are nonlinear travelling waves with a central core that demonstrates classical solitary wave behaviour, as well as non-decaying oscillations that continue away from the core indefinitely in one or both directions, and have amplitude that is exponentially small in some asymptotic parameter. This is distinct from classical solitary waves, such as those exhibited by the Korteweg-de Vries equation (KdV) discussed below, which are localized in a frame that moves with the wave core. Examples of solitary waves and GSW are displayed in Figure \ref{F:GSW}. 

These GSW, also known in the literature as `nanoptera', have been studied in a physical contexts, particularly arising in fluid dynamics, including solitary water waves with surface tension. This problem has been studied analytically in \cite{Grimshaw2, Grimshaw1, Pomeau1, Trinh5}, typically using a fifth-order Kortweg de Vries model. Numerical studies illustrating GSW in gravity-capillary wave systems include \cite{Champneys1,Clamond1,VandenBroeck6}, and the existence of these waves was rigorously proven by \cite{Beale1, Iooss2, Sun1}. 


GSW have also more recently been studied in discrete problems, including both difference and differential-difference equations. Systems that have been shown to demonstrate this behaviour include discrete nonlinear Schr\"odinger lattices \cite{Alfimov2,Melvin1,Melvin2,Oxtoby1}, Fermi--Pasta--Ulan--Tsingou systems \cite{Faver1,Hoffman1}, and Toda chains \cite{Lustri5,Vainchtein1}, the discrete Klein-Gordon equation \cite{Alfimov1}, the Frenkel-Kontorova dislocation model \cite{King4}, and nonlinear chains of oscillators \cite{Iooss1}. Given that there are many difference equations which produce GSW, it seems natural to expect that there are also lattice equations which produce GSW solutions.

Typically, GSW occur in singular perturbations of systems that already contain solitary wave solutions. For example, generalized solitary water waves with surface tension are governed by a fifth-order perturbation to the classical KdV equation. In their studies of this equation, Pomeau \textit{et al.} \cite{Pomeau1} and Grimshaw \& Joshi \cite{Grimshaw2} used Borel summation of the divergent asymptotic series to determine the behaviour of amplitude of the oscillations present on either side of the wave core. It is reasonable to ask if \textit{all} singular perturbations of the KdV equation produce GSW, and if not, what conditions are required to produce this phenomenon. This question was addressed by Pomeau \textit{et al.} \cite{Pomeau1}, who noted that while the fifth-order KdV does produce GSW, this is not necessarily true of the corresponding seventh-order KdV. 

The seventh-order KdV equation (7KdV) discussed in \cite{Pomeau1} is given by
\begin{equation}
E_{\epsilon}\left\{u\right\}:= \lambda \epsilon^4 \pdiff{^7 u}{x^7} + \epsilon^2 \pdiff{^5u}{x^5} + \pdiff{^3 u}{x^3} + 6 u \pdiff{u}{x} +\pdiff{u}{t} = 0.\label{2.7KdV}
\end{equation}
This is not a water-wave equation, and therefore does not have the same physical relevance as the fifth-order KdV equation. Instead, this was viewed as a testing ground for understanding the general behaviour of singular perturbations to the undisturbed KdV equation, given by
\begin{equation}
E_{0}\left\{u\right\}:=\pdiff{^3 u}{x^3} + 6 u \pdiff{u}{x} + \pdiff{u}{t} = 0.\label{0.kdv}
\end{equation}
Pomeau \textit{et al.} \cite{Pomeau1} applied Borel summation methods to show that 7KdV only possesses GSW solutions for $\lambda \leq \tfrac{1}{4}$. If $\lambda$ is greater than critical value, any oscillations are damped in the far field, leading to a localized classical solitary wave solution. In later work, Grimshaw \textit{et al.} \cite{Grimshaw6} showed that similar results do hold for the corresponding fifth-order KdV equation
\begin{equation}
\kappa \epsilon^2 \pdiff{^5u}{x^5} + \pdiff{^3 u}{x^3} + 6 u \pdiff{u}{x} +\pdiff{u}{t} = 0,\label{0.kdv5}
\end{equation}
finding that the critical value of $\kappa$ is zero.

In subsequent work, Grimshaw \& Joshi \cite{Grimshaw1}, showed that the results of Pomeau \textit{et al.} for the fifth-order KdV equation could be explained by studying this equation using exponential asymptotic techniques. This was further confirmed in later studies \cite{Grimshaw2,Trinh5}. They determined that the GSW could be understood be consider the behaviour of special curves in the solution domain, known as `Stokes lines', across which exponentially small behaviour in the solution appears. 

In this paper, we will show that the results of \cite{Pomeau1} for 7KdV can also be explained by using direct exponential asymptotic methods. Using exponential asymptotics, we will determine the location of Stokes lines in the solution, and we will directly compute the exponentially-small oscillations that are switched as these Stokes lines are crossed. From these computations, we will see that these oscillations decay away from the core of the wave when $\lambda > \tfrac{1}{4}$ producing localized solitary waves, as identified by \cite{Pomeau1}. While similar in some respects to the analysis of \cite{Pomeau1}, this method explicitly demonstrates the important role played by Stokes Phenomenon in the asymptotic solution. Furthermore, using this Stokes line analysis technique, we will extend this result and determine corresponding critical values for a full KdV hierarchy, and subsequently apply an adapted form this this methodology to study GSW in a lattice KdV equation.

For more detailed descriptions of the foundations of exponential asymptotics, see \cite{Berry3, Berry4, Berry5}. For more details on the exponential asymptotic techniques used in this paper, see \cite{Chapman1, Daalhuis1}.

\section{Seventh-Order KdV}

In this section, we study the exponentially small oscillations present in travelling wave solutions to the seventh-order KdV equation \eqref{2.7KdV}. Using exponential asymptotics, we will show that these oscillations are switched on across special curves known as Stokes lines, and recover the condition that GSW solutions appear for $\lambda \leq \tfrac{1}{4}$ \cite{Pomeau1}.

\subsection{Formulation and Series Expansion}



Consider 7KdV, given by $E_{\epsilon}\left\{ u\right\} = 0$ in \eqref{2.7KdV}. We expand $u(x,t)$ as a power series in $\epsilon^2$, 
\begin{equation}
u(x,t) \sim \sum_{j=0}^{\infty} \epsilon^{2j} u_j(x,t).\label{2.series}
\end{equation}

Applying this expansion to 7KdV and matching at $\mathcal{O}(1)$ as $\epsilon\rightarrow 0$ gives $E_0\left\{u_0\right\} = 0$, showing that $u_0$ satisfies the KdV equation \eqref{0.kdv}. We select $u_0(x,t)$ to be the one-soliton solution to the KdV equation, given by
\begin{equation}
u_0(x,t) = \tfrac{1}{2}c \,\mathrm{sech}^2\left[\tfrac{1}{2}\sqrt{c}\,(x - ct)\right],\label{2.u0}
\end{equation}
where $c$ is an arbitrary choice of the soliton speed. \textcolor{black}{We will also include the condition that the full asymptotic solution must be symmetric about the wave core, located at $x = ct$. }

\textcolor{black}{At this point, we note that previous analyses of the KdV equation, as well as higher-order KdV equations, frequently introduce a moving reference frame (say, $\nu = x-ct$) in order to express the system as an ordinary differential equation, as well as impose the symmetry condition about $\eta = 0$. This approach would work for the present analysis; however, we would like to allow for the possibility of leading-order KdV solutions that are more complicated than a single travelling wave, such as interacting solitons, or solutions with arbitrary initial conditions. While we do not consider such systems in the present study, they present interesting mathematical challenges that could be considered in further work. We discuss this possibility in more detail in Section \ref{S:Discuss}.}

Analytically continuing the solution in the $x$-plane shows that $u_0(x,t)$ has singularities at $x_s =  ct + {\mathrm{i} \pi M}/{\sqrt{c}}$, where $M\in 2 \mathbb{Z}+1$. In the neighbourhood of these singularities, the leading-order behaviour is given by
\begin{equation}
u_0(x,t) \sim -\frac{2}{(x - x_{s})^2} \qquad \mathrm{as} \qquad x \rightarrow x_{s}.\label{2.singlocal}
\end{equation}
The singularities that will dominate the solution behaviour for real (physical) values of $x$ are those closest to the real $x$ axis, associated with $M = \pm 1$. In subsequent analysis, we will denote the location of these singularities as $x_+$ and $x_-$.

 To find subsequent terms in the series, we must match apply the series \eqref{2.series} to 7KdV and match at higher orders of $\epsilon$. Matching at $\mathcal{O}(\epsilon^{2k})$ for $k \geq 2$ gives the general recurrence equation
 \begin{equation}
\lambda \pdiff{^7 u_{k-2}}{x^7} + \pdiff{^5u_{k-1}}{x^5} + \pdiff{^3 u_k}{x^3} + 6 \sum_{r=0}^{k} u_r \pdiff{u_{k-r}}{x} + \pdiff{u_k}{t} = 0,\qquad k \geq 2.\label{2.recur}
\end{equation}

Repeatedly applying this recursion relation produces terms in the series \eqref{2.series}; however, this will not capture the far-field oscillations, as they are exponentially small in the limit $\epsilon \rightarrow 0$ and hence beyond the reach of the algebraic series expression. However, this recurrence relation permits us to obtain the form of $u_j$ in the limit that $j \rightarrow \infty$, known as the late-order series behaviour.

\subsection{Late-Order Asymptotic Terms}\label{S:LOT2}

Following \cite{Dingle1,Daalhuis1}, we observe that this is a singularly-perturbed problem with singularities in the analytically continued leading-order behaviour. This series will therefore diverge in a predictable factorial-over-power fashion, as the late-order behaviour is dominated by the results of repeatedly differentiating the singular term \cite{Dingle1}. We therefore follow \cite{Chapman1} and propose an ansatz for the late-order terms of the form
\begin{equation}
u_j \sim \frac{F(x,t) \Gamma(2j + \gamma)}{\chi(x,t)^{2j+\gamma}},\qquad \mathrm{as} \qquad j \rightarrow \infty.\label{2.ansatz}
\end{equation}
Substituting this expression into the recurrence relation \eqref{2.recur} and taking the leading order as $k \rightarrow \infty$ gives
\begin{align}
\nonumber -\frac{F[\lambda\chi_x^{7} + \chi_x^5 + \chi_x^3]\Gamma(2k+ 3 + \gamma)}{\chi^{2k + 3 + \gamma}}+ &\frac{F_x[7\lambda\chi_x^{6} + 5\chi_x^4 + 3 \chi_x^2]\Gamma(2k+ 2 + \gamma)}{\chi^{2k + 2 + \gamma}}\\
&+ \frac{3 F \chi_{xx} [6 \lambda\chi_x^5 + 4 \chi_x^3 + 2 \chi_x]\Gamma(2k+ 2 + \gamma)}{\chi^{2k + 2 + \gamma}} + \ldots = 0.
\label{2.late recur}
\end{align}
where the terms omitted are $\mathcal{O}(u_{k+1/2})$ in the limit as $k \rightarrow \infty$. 

Matching at $\mathcal{O}(u_{k + 3/2})$ as $j\rightarrow \infty$ gives $\lambda\chi_x^7 + \chi_x^5 + \chi_x^3= 0$. Noting that $\chi_x$ cannot be zero, we simplify this to give the singulant equation
\begin{equation}
\lambda\chi_x^4 + \chi_x^2 + 1= 0.\label{2.singulant0}
\end{equation}
This has four solutions, 
\begin{equation}
\chi_x = \pm\mathrm{i} \sqrt{\tfrac{1}{2\lambda} \pm \sqrt{\tfrac{1}{4\lambda}-1}},
\end{equation}
where the signs are chosen independently.

If $\lambda > \tfrac{1}{4}$, the solutions to this equation are complex, with nonzero real and imaginary parts. We denote the solution with positive real and imaginary part as $\chi_x = \alpha$. The four solutions are given by $\chi_x = \pm\alpha$ and $\pm\overline{\alpha}$, where the bar corresponds to complex conjugation. 

If $\lambda \leq \tfrac{1}{4}$, the solutions are imaginary conjugate pairs. We denote the solution with smallest positive imaginary part as $\chi_x = \mathrm{i} \beta$, with $\beta \in \mathbb{R}^+$.  

We first consider the case $\lambda > \tfrac{1}{4}$. Recalling that the singulant must be zero at the location of the leading-order singularities, denoted $x_s$, we solve \eqref{2.singulant0} to give
\begin{equation}
\chi _1= \alpha (x - x_s), \qquad \chi_2 = -\overline{\alpha} (x - x_s),\qquad \chi_3 = -\alpha (x - x_s),\qquad \chi_4 = \overline{\alpha} (x - x_s),
\end{equation}
The two singularities, located at $x_s = x_+$ and $x_s = x_-$ therefore each have four associated late-order contributions, corresponding to the four singulants. 

If we perform a full analysis on all eight contributions, we find that only four produce Stokes switching that affects real values of $x$. We recall that Stokes switching only occurs in asymptotic contributions that are exponentially small. In subsequent analysis, we will find that the late-order contributions have a form given by \eqref{2.RNWKB}. From this form, we see that the remainder contribution is only exponentially small for real $x$ if $\mathrm{Re}(\chi) > 0$ on the real axis, and therefore Stokes switching may only affect these terms. Consequently, we restrict our attention to the four terms that satisfy this condition, associated with
%
\begin{equation}
\chi _1= \alpha (x - x_+), \qquad \chi_2 = -\overline{\alpha} (x - x_+),\qquad \chi_3 = \overline{\alpha} (x - x_-),\qquad \chi_4 = -{\alpha} (x - x_-).\label{2.singulant}
\end{equation}
The remaining analysis will be performed in detail for the contribution associated with $\chi_1$, with corresponding results stated for the remaining singulants. To determine the prefactor associated with $\chi_1$, we return to \eqref{2.late recur} and match at $\mathcal{O}(u_{k+1})$, noting that $\chi_{xx} = 0$, we find that $F_x = 0$. Consequently, $F$ is independent of $x$.

We require that the singularity strength of the ansatz \eqref{2.ansatz} in the limit that $x$ approaches the singularity must be consistent with the local leading-order singular strength in \eqref{2.singlocal}. The inner limit of this outer behaviour is given by
\begin{equation}
u_j(x,t) \sim \frac{F(t)\Gamma(2j+\gamma)}{[\alpha (x-x_+)]^{2j+\gamma}} \qquad \mathrm{as} \qquad x \rightarrow x_+ \quad \mathrm{and} \quad j \rightarrow \infty.\label{2.ansatzinner}
\end{equation}
From the recurrence relation \eqref{2.recur}, it is apparent that strength of the singularity at $x = x_+$ increases by two at each order, and we therefore require $\gamma = 2$ in order for the singularity strength of this term to be consistent with the leading-order singularity behaviour given in \eqref{2.singlocal}. 

To determine the value of $F$, we now match the behaviour of the late-order terms with a local expansion in the neighbourhood of the singularity at $x_s$. This inner analysis is performed in Appendix \ref{S.Inner0}, and shows that the prefactors for each of the contributions are constant. The constant associated with $\chi_1$ is denoted by $F(x,t) = \Lambda$. In Appendix \ref{S.Inner0}, we perform an example calculation with $\lambda = 1$ that shows $\Lambda \approx 0.711 + 0.694\mathrm{i}$. We subsequently find that the corresponding constants associated with $\chi_2$ and $\chi_3$ are equal to $\overline{\Lambda}$, while the constant associated with $\chi_4$ is also equal to $\Lambda$.

The late order terms responsible for Stokes switching in the solution are hence given by
\begin{equation}
u_j \sim \frac{\Lambda\Gamma(2j + 2)}{[\alpha(x - x_+)]^{2j+2}} + \frac{\overline{\Lambda}\Gamma(2j + 2)}{[-\overline{\alpha}(x - x_+)]^{2j+2}} +\frac{ \overline{\Lambda}\Gamma(2j + 2)}{[\overline{\alpha} (x - x_-)]^{2j+2}}+ \frac{ {\Lambda}\Gamma(2j + 2)}{[-\alpha (x - x_-)]^{2j+2}},\label{2.LOT}
\end{equation}
as $j\rightarrow \infty$ where $x_\pm = ct\pm\mathrm{i} \pi /\sqrt{c} $.

A similar analysis may be applied to the case $\lambda \leq \tfrac{1}{4}$. In this case, we need only consider the singulants with smallest absolute value on the real axis, which will dominate the late-order behaviour. We therefore have only two relevant contributions, $\chi_x = \pm\mathrm{i}\beta$. The form of the late-order terms is given by 
\begin{equation}
u_j \sim \frac{\Lambda\Gamma(2j + 2)}{[\mathrm{i}\beta(x - x_+)]^{2j+2}} + \frac{ {\Lambda}\Gamma(2j + 2)}{[-\mathrm{i}\beta (x - x_-)]^{2j+2}},\label{2.LOTb}
\end{equation}
The inner analysis for this problem required to determine the constant value of $\Lambda$ is significantly simpler than the $\lambda > \tfrac{1}{4}$ case, as each singularity corresponds to only one late-order contribution in \eqref{2.LOTb}. The differences in the inner analysis are discussed in Appendix \ref{S.Inner0}, in which we also show that $\Lambda$ is always real for this case. We demonstrate an example inner analysis for $\lambda = \tfrac{1}{8}$, finding that $\Lambda \approx -11.70$.

\subsection{Exponential Asymptotics}

In this section, we perform straightforward application of the matched asymptotic expansion technique described in \cite{Daalhuis1} in order to determine the amplitude of the exponentially small oscillations present in the solution. We therefore omit several technical computations, and direct the reader to this reference for further details.

%
%

To perform an exponential asymptotic analysis on the late-order terms from \eqref{2.LOT} and \eqref{2.LOTb}, we must truncate the asymptotic series optimally, ensuring that the remainder is exponentially small \cite{Berry4, Berry5,Boyd1}. To find the optimal truncation point, which we denote as $N$, we follow the commonly-used heuristic described by \cite{Boyd1}, in which the series is truncated at its smallest term.  The resultant expression gives $N \sim |\chi|/2\epsilon$; we therefore set $N = |\chi|/2\epsilon + \omega$, where $0 \leq \omega < 1$ is selected so that $N \in \mathbb{Z}$. We note that $N \rightarrow \infty$ in the limit that $\epsilon \rightarrow 0$, justifying the assumption that $N$ is large in the asymptotic limit.

We truncate the series \eqref{1.series} at the optimal truncation point to give
\begin{equation}
u(x,t) = \sum_{j = 0}^{N-1} \epsilon^{2j} u_j + R(x,t),
\end{equation}
where $R$ is the truncation remainder. 

We apply this expression to \eqref{2.7KdV}, and use \eqref{2.recur} to eliminate series terms, leaving
\begin{align}
\nonumber \lambda\epsilon^4 \pdiff{^7 R}{x^7} + \epsilon^2 \pdiff{^5 R}{x^5}  + \pdiff{^3 R}{x^3} & +  6 R \sum_{j=0}^{N-1}\epsilon^{2j} \pdiff{u_j}{x} + 6 \pdiff{ R}{x}  \sum_{j=0}^{N-1}\epsilon^{2j} u_j + \pdiff{R}{t}  \\
&= - \lambda\epsilon^{2N+2}\pdiff{^7u_{N-1}}{x^7}- \lambda\epsilon^{2N}\pdiff{^7u_{N-2}}{x^7} - \epsilon^{2N}\pdiff{^5u_{N-1}}{x^5}.
\end{align}
Keeping only the largest terms in the limit that $\epsilon \rightarrow 0$ and $N \rightarrow \infty$, and applying \eqref{2.recur} to simplify the right-hand side of the expression, we obtain
\begin{align}
\lambda \epsilon^4 \pdiff{^7 R}{x^7} + \epsilon^2 \pdiff{^5 R}{x^5}  + \pdiff{^3 R}{x^3} + \ldots =  \epsilon^{2N}\pdiff{^3 u_N}{x^3} + \ldots,\label{2.R}
\end{align}
where the omitted terms do not contribute to the dominant behaviour as $\epsilon \rightarrow 0$.

The right-hand side only contributes to the remainder behaviour in the neighbourhood of the Stokes line.  We can therefore determine the behaviour away from the Stokes line by solving the homogeneous problem for $R$. Using a WKB (or Liouville-Green) ansatz, $u \sim f(x,t) \e^{-g(x,t)/\epsilon}$, we see that this is satisfied by $g = \chi$. Consequently, we select the form of the remainder in the limit that $\epsilon \rightarrow 0$ as
\begin{equation}\label{2.RNWKB}
R \sim \mathcal{S} \e^{-\chi/\epsilon},
\end{equation}
where $\mathcal{S}(x,t)$ is the Stokes multiplier that captures the variation in the neighbourhood of the Stokes line. 

Applying this expression to \eqref{2.R} and applying the late-order ansatz \eqref{2.LOT} gives
\begin{align}
\mu \mathcal{S}_x \e^{-\chi/\epsilon} \sim -\frac{\epsilon^{2N+2}\Lambda (\chi_x)^3 \Gamma(2N+5)}{\chi^{2N+5}} \qquad \mathrm{as} \qquad \epsilon \rightarrow 0,\label{2.Seq1}
\end{align}
where $\mu = 7\lambda \chi_x^6 + 5\chi_x^4 + 3\chi_x^2$, recalling that the limit  $\epsilon \rightarrow 0$  corresponds to  $N \rightarrow \infty$. 

We now set the optimal truncation point, $N = |\chi|/2\epsilon + \omega$, and express the singulant using polar terms so that $\chi = \rho \e^{\mathrm{i}\theta}$. We note that the right-hand side is exponentially small except on $\theta = 0$, which is the Stokes line location. This corresponds to $\mathrm{Re}(\chi) > 0$ and $\mathrm{Im}(\chi) = 0$, or the straight line extending downwards form the singularity along $\mathrm{Re}(x) = ct$.

We define a set of inner coordinates to consider \eqref{2.Seq1} in a neighbourhood of width $\mathcal{O}(\sqrt{\eps})$ near the Stokes line. This process can be seen in more detail in \cite{Daalhuis1}. Solving the inner equation and converting back to outer coordinates gives
%
%
\begin{equation}
\mathcal{S} = -\frac{\mathrm{i} \Lambda \chi_x^2 \sqrt{2\pi \rho}}{\mu \epsilon^2 }\int_{-\infty}^{\sqrt{\rho}\theta/\eps} \e^{-s^2/2}\mathrm{d} s + \mathcal{S}_0,\label{2.Sint}
\end{equation}
where $\mathcal{S}_0$ is a constant of integration. Consequently, the jump in $\mathcal{S}$ as the Stokes line is crossed from left to right is given by
\begin{equation}
\left[\mathcal{S}\right]_-^+ \sim -\frac{2\pi \mathrm{i} \chi_x^2 \Lambda}{\mu \epsilon^2}.\label{2.sjump}
\end{equation}
At this point, the exponential asymptotic analysis is valid for general $\chi$, and may therefore be applied to both \eqref{2.LOT} and \eqref{2.LOTb}. In order to analyse the jump across the Stokes line, it is useful to consider $\lambda > \tfrac{1}{4}$ and $\lambda \leq \tfrac{1}{4}$ seperately. We denote these as Case A and Case B solutions respectively.

\subsubsection{Case A}

\begin{figure}
\begin{center}
\includegraphics{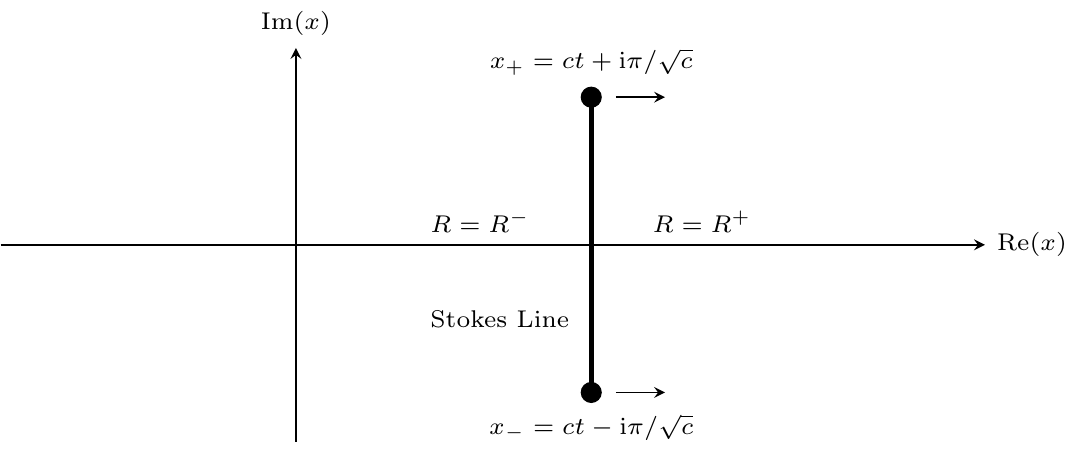}
%
%
%
%
%
%
%

\end{center}
\caption{Stokes line in $u(x,t)$ across which the exponentially small contributions to the solution switch for both cases of the 7KdV equation \eqref{2.7KdV}, and the lKdV equation \eqref{1.fdkdv}. The singularities are denoted by circles, and the Stokes line is depicted as a thick line. On the left of the Stokes line, the exponential contribution $R$ is given by $R^-$, and on the right of the Stokes line, the contribution is given by $R^+$.}\label{F.7KdVstokes}
\end{figure}

Using the remainder ansatz \eqref{2.RNWKB} and the value of $\chi_1$, the jump in the Stokes multiplier \eqref{2.sjump} corresponds to a jump in the remainder of
\begin{equation}
\left[R_1\right]_-^+ \sim  -\frac{2\pi \mathrm{i} \alpha^2 \Lambda}{\mu \epsilon^2}\e^{-\alpha(x-ct - \mathrm{i}\pi/\sqrt{c})/\epsilon},
\end{equation}
where we have denoted this as $R_1$ to indicate that it is the contribution associated with $\chi_1$. We require that the oscillations are exponentially small in the far field; however, this contribution becomes exponentially large for $x < ct-\pi\sqrt{3/c}$. This necessitates that the exponential contribution be inactive in this region. Consequently, this contribution is inactive on the left-hand side of the Stokes line, and switched on as the Stokes line at $x = ct$ is crossed from left to right. Therefore, the remainder $R_1$ switches from $R_1^-$ to $R_1^+$ as the Stokes line is crossed from left to right, where 
\begin{equation}
R_1^- = 0,\qquad R_1^+ \sim  -\frac{2\pi \mathrm{i} \alpha^2 \Lambda}{\mu \epsilon^2}\e^{-\alpha(x-ct - \mathrm{i}\pi/\sqrt{c})/\epsilon}.\label{2.R1}
\end{equation}
This Stokes structure is illustrated in Figure \ref{F.7KdVstokes}.

The remaining contributions can be calculated using nearly-identical analyses. The contribution $R_3$ gives a complex conjugate expression to \eqref{2.R1}. The contributions $R_2$ and $R_4$ are complex conjugate quantities that are exponentially large in the region $x > ct + \pi\sqrt{3/c}$. These quantities are therefore present to the left of the Stokes line, and switched off as the Stokes line is crossed from left to right. $R_2$ and $R_4$ are, in fact, a reflection of $R_1$ and $R_3$ respectively about the point $x = ct$; consequently, the system is symmetric about this point. The full remainder $R$ is given by
\begin{equation}
R^+ \sim -\frac{4\pi}{\mu \epsilon^2}\e^{-[\alpha_r(x-ct) + \alpha_i\pi/\sqrt{c}]/\epsilon} \mathrm{Re}\left[ \alpha^2 \Lambda\e^{\mathrm{i} [\alpha_i(x-ct) -\alpha_r\pi/\sqrt{c}]/\epsilon} \right],\label{2.R+}
\end{equation}
for $x > c t + \delta$, and
\begin{equation}
R^+ \sim -\frac{4\pi}{\mu \epsilon^2}\e^{-[\alpha_r(ct-x) + \alpha_i\pi/\sqrt{c}]/\epsilon} \mathrm{Re}\left[ \alpha^2 \Lambda\e^{\mathrm{i} [\alpha_i(ct-x) -\alpha_r\pi/\sqrt{c}]/\epsilon} \right] \label{2.R-}
\end{equation}
for $x < c t - \delta$, where $\alpha = \alpha_r + \mathrm{i} \alpha_i$, and $\delta  = \mathcal{O}(\epsilon^{1/2})$ denotes the inner region that contains the smooth transition between the two remainder expressions.

\begin{figure}
\begin{center}
%
%
%
%
%
\includegraphics{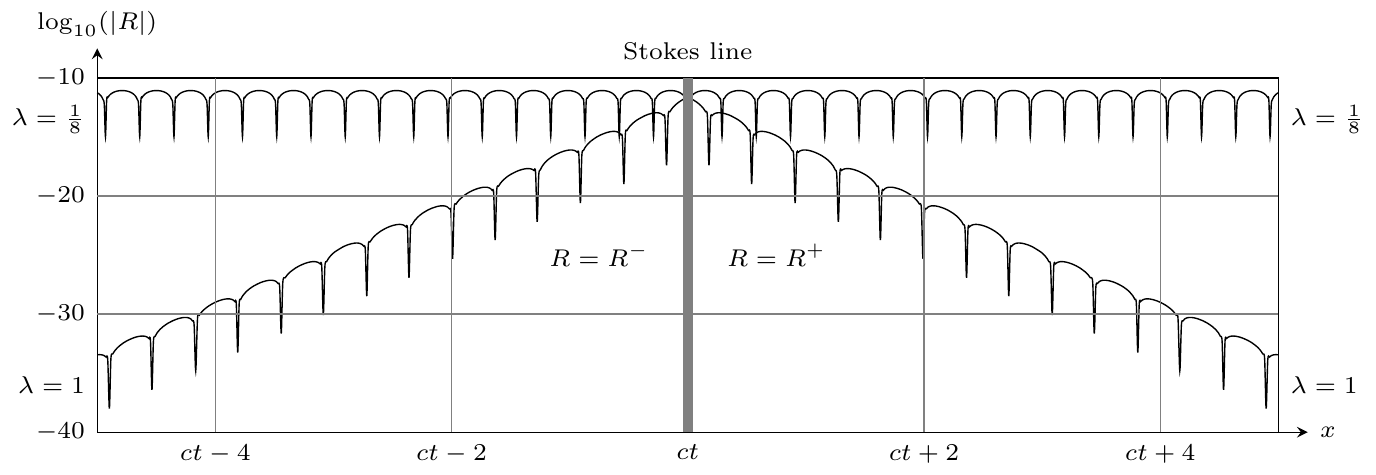}

\end{center}
\caption{Logarithmic plot of $|R|$ for $\epsilon = 0.1$ and $c = 1$, with $\lambda = 1$ and $\lambda = {1}/{8}$. On the left-hand side of the Stokes line, the remainder is given by $R^-$ \eqref{2.R-}, and on the right of the Stokes line, the remainder is given by $R^+$ \eqref{2.R+}. For $\lambda = 1$, the the remainder is given in \eqref{2.RcaseB} and has constant, non-decaying amplitude.} \label{F.7KdVR}
\end{figure}

The remainder expressions in \eqref{2.R+} and \eqref{2.R-} contain rapid, exponentially small oscillations. The amplitude of these oscillations decays as $x-ct \rightarrow \pm\infty$. This solution to \eqref{2.7KdV} therefore does not produce GSW, in contrast to the singularly-perturbed fifth-order KdV equation. Instead, the solitary wave in 7KdV is exponentially localized in space. The decaying oscillations are shown in Figure \ref{F.7KdVR}.

\subsubsection{Case B}\label{S.CaseB}

In this case, we instead find that the jump in the exponential behaviour is given by
\begin{equation}
\left[R_1\right]_-^+ \sim  \frac{2\pi \mathrm{i} \beta^2 \Lambda}{\mu \epsilon^2}\e^{-\mathrm{i}\beta(x-ct-\mathrm{i}\pi/\sqrt{c})/\epsilon},
\end{equation}
The remainder term clearly does not grow exponentially on either side of the Stokes line, and therefore we are free to select $\mathcal{S}_0$ in \eqref{2.Sint} arbitrarily. In order for the solution to be symmetric about the point $x = ct$, we must select $\mathcal{S}_0$ so that $\mathcal{S} = 0$ on the Stokes line. The remainder term switches rapidly from $\mathcal{S}^-$ to $\mathcal{S}^+$ as the Stokes line is crossed, where
\begin{equation}
\mathcal{S}^- \sim -\frac{\pi \beta^2 \Lambda}{\mu \epsilon^2},\qquad \mathcal{S}^+ \sim  \frac{\pi \beta^2 \Lambda}{\mu \epsilon^2}.\label{2.Sswitch}
\end{equation}
We perform a similar analysis on the late-order terms associated with $\chi_2$, and find the resultant contribution is the complex conjugate of $R_1$. From the switching information given in \eqref{2.Sswitch}, and knowing that $R \sim \mathcal{S}\e^{-\chi/\epsilon}$, we find that the combined exponentially small remainder term $R$ switches from $R^-$ to $R^+$ as the Stokes line at $\mathrm{Re}(x) = ct$ is crossed from left to right, where these quantities are given by
\begin{equation}
R^- \sim -\frac{2\pi\beta^2\Lambda}{\mu \epsilon^2}\e^{-\beta\pi/\epsilon\sqrt{c}}\sin\left[\frac{\beta}{\epsilon}(x-ct)\right],\qquad
R^+ \sim \frac{2\pi\beta^2\Lambda}{\mu \epsilon^2}\e^{-\beta\pi/\epsilon\sqrt{c}}\sin\left[\frac{\beta}{\epsilon}(x-ct)\right].\label{2.RcaseB}
\end{equation}
These oscillations are symmetric about $x = ct$, as required. Consequently, Figure \ref{F.7KdVstokes} also shows the Stokes structure for the $\lambda < \tfrac{1}{4}$ case, where the remainder terms are non-decaying oscillatory functions given in \eqref{2.RcaseB}. 

We see that, as first noted in \cite{Pomeau1}, 7KdV only produces GSW if $\lambda \leq \tfrac{1}{4}$, and instead has exponentially localized solutions if $\lambda > \tfrac{1}{4}$. The behaviour of $R$ for $\lambda = \tfrac{1}{8}$ is illustrated in Figure \ref{F.7KdVR}. It is clear that these contributions do not decay in the far field, unlike the example shown for $\lambda = 1$.

\textcolor{black}{Finally, we note that in Case A $(\lambda > \tfrac{1}{4}$), the sum of the four exponentially small components satisfied the symmetry condition about $x = ct$ that was imposed on the solution, as each contribution could only be active on one side of the Stokes line. The exponentially-small contribution is therefore unique and symmetric. In Case B ($\lambda \leq \tfrac{1}{4}$), the exponentially small oscillations could be active on both sides of the Stokes line, and we instead enforced the symmetry requirement by selecting the value $\mathcal{S}_0$ appropriately. This allows for the possibility that we could make a different choice, such as selecting $\mathcal{S}_0$ to restrict the oscillations to lie on only one side of the central wave core (see, for example, \cite{Lustri5}), or in some other asymmetric configuration. }

\subsection{Comparison of Results}

\begin{figure}
\centering
%
%
%
%
%
%
\includegraphics{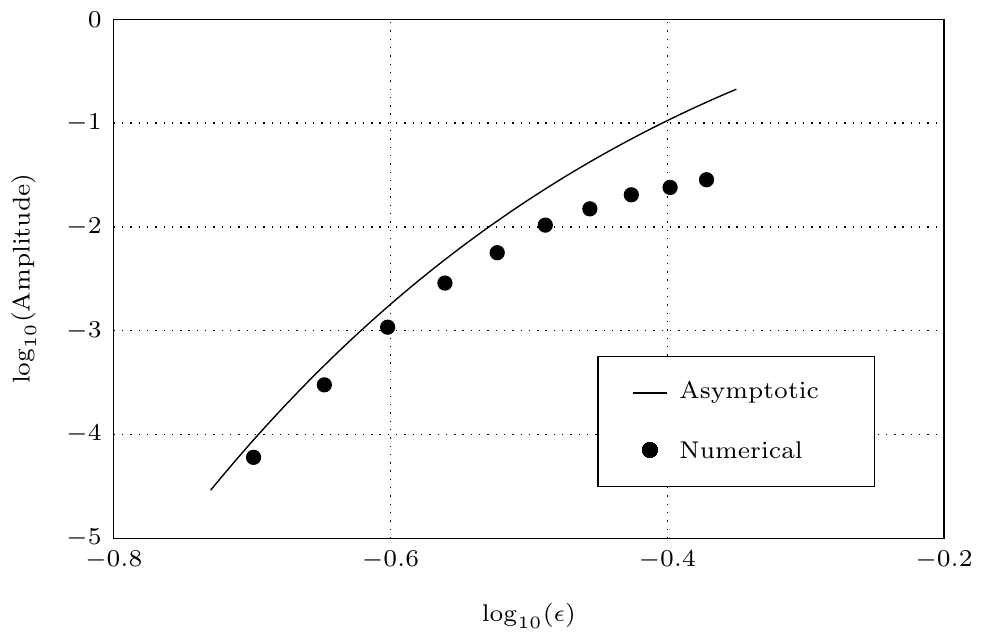}

\caption{Comparison between far-field oscillation amplitude predicted by exponential asymptotics \eqref{2.RcaseB}, shown as a solid line, and the amplitude calculated using a numerical split-step method, shown as circles. The results were calculated for $\lambda = \tfrac{1}{8}$ and $c = 1$. As expected, the asymptotic analysis becomes more accurate as $\epsilon \rightarrow 0$.}\label{F:Comparison}
\end{figure}

\textcolor{black}{In Figure \ref{F:Comparison}, we compare the asymptotic prediction of the far-field amplitude of the GSW with velocity $c = 1$ in a system with $\lambda = \tfrac{1}{8}$ with the corresponding numerically-calculated far-field amplitude. The numerical amplitudes were obtained by adapting the split-step method for solving the KdV equation, introduced in \cite{Tappert1}. This commonly-used method involves splitting the time-stepping procedure into a linear part and a nonlinear part; the linear step is calculated using the fast Fourier transform in the frequency domain, while the nonlinear step is computed directly in the time domain. For more details on split-step methods see, for example, \cite{Tappert1}.}

\textcolor{black}{We computed the solution of the 7KdV equation on a large periodic domain using $u_0(x,0)$ from \eqref{2.u0} as the initial condition. As the system evolved in time, exponentially small oscillations spread from the right-hand side of the wave core. The total time was taken to be sufficiently large for the oscillation amplitude to reach a steady value. For the values considered in Figure \ref{F:Comparison}, the system was computed over $0 \leq t \leq 3$, and $-50 \leq x \leq 50$. A similar numerical approach was used to calculate the amplitude of GSW oscillations in the fifth-order KdV equation in \cite{Benilov1}. }

\textcolor{black}{Importantly, using this method produced solutions in which the oscillations appear one one side of the central wave core, producing a one-sided GSW solution similar to the illustration in Figure \ref{F:GSW} (c). From the jump condition \eqref{2.Sswitch}, we know that the amplitude of the symmetric GSW solution is obtained by halving the far-field amplitude of the one-sided GSW solution. A comparison between the results of this computation and the  asymptotic prediction from \eqref{2.RcaseB} is depicted in Figure \ref{F:Comparison}. It is apparent from this figure that the asymptotic analysis becomes more accurate as $\epsilon \rightarrow 0$, providing numerical support for the asymptotic approximation.}

\subsection{Higher-order KdV equations}\label{S:Higher}

The decay of the exponentially small oscillations in 7KdV for $\lambda > \tfrac{1}{4}$ could have been predicted  directly from the form of the singulant \eqref{2.singulant0}. Knowing that the form of the exponentially small remainder is generally \eqref{2.RNWKB}, it is clear that non-decaying oscillations correspond to imaginary values of $\chi_x$. Any real component of $\chi_x$ will cause the remainder to grow or decay exponentially in space, violating the definition of a GSW.

For example, we consider a general class of singularly-perturbed KdV equations of order $n = 2k+3$ for $k \geq 1$. This class contains \eqref{2.7KdV}, and is given by
\begin{equation}
\lambda\epsilon^{2k} \pdiff{^{2k+3}u}{x^{2k+3}} + \sum_{r=0}^{k-1} \epsilon^{2r} \pdiff{^{2r+3} u}{x^{2r+3}}+ 6 u\pdiff{u}{x} +\pdiff{u}{t} = 0,\label{2.nkdv2}
\end{equation}
with $\lambda \in \mathbb{R}$ and $\lambda \neq 0$. If we expand this as a power series using \eqref{2.series}, the behaviour of $u_0$ is always governed by the KdV equation, and therefore permits solutions of the form \eqref{2.u0}. Obtaining the late-order terms using an ansatz, as in Section \ref{S:LOT2}, we find that the singulant satisfies
\begin{equation}
\lambda \chi^{2k}_x + \sum_{r=0}^{k-1} \chi_x^{2r} = 0,\qquad \lambda \neq 0.\label{2.nsing2}
\end{equation}
The roots of this equation determine which solitary wave solutions contain exponentially small non-decaying oscillations. We identify the presence of GSW by finding imaginary solutions for $\chi_x$ for particular combinations of $\lambda$ and $k$.

If $k$ is odd, \eqref{2.nsing2} has imaginary solutions for $\chi_x$ if $\lambda > 0$, but no imaginary solutions if  $\lambda < 0$. Therefore travelling wave solutions for $\lambda > 0$ are GSW, while for $\lambda < 0$, all oscillations decay in the far field, meaning that the solitary wave is exponentially localized in space. This is consistent with the critical value of $\lambda = 0$ obtained in \cite{Grimshaw6} for the fifth-order KdV equation, which corresponds to $k$ = 1.

\begin{figure}
\centering
\includegraphics{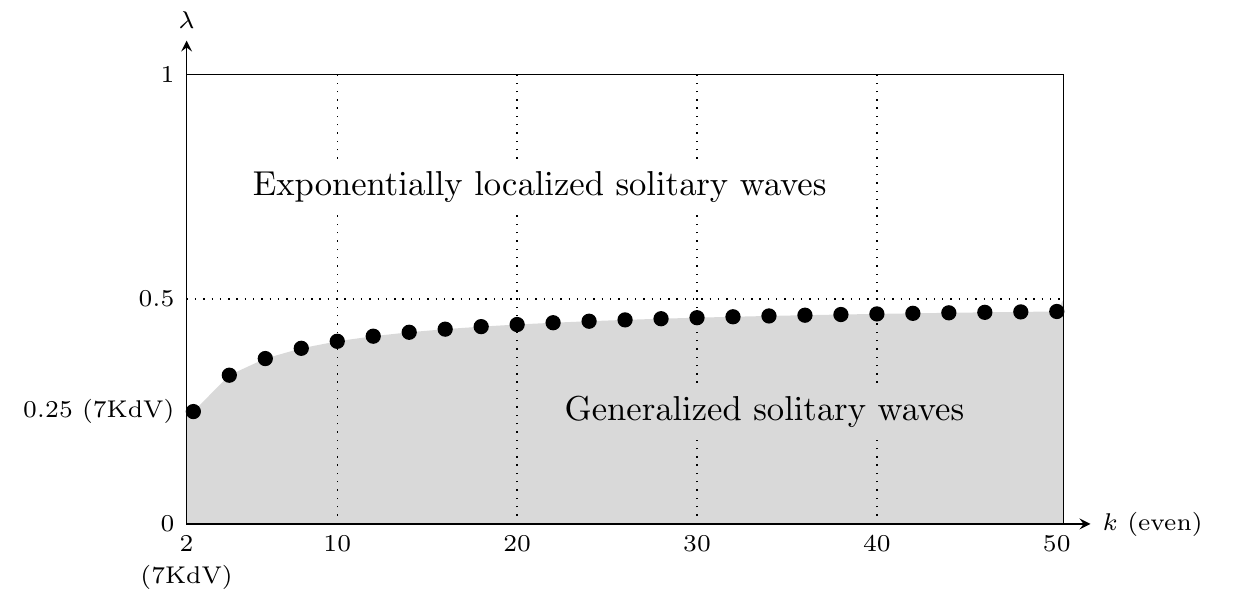}

\caption{Values of $\lambda_{\mathrm{crit}}$ for a range of even $k$ values, denoted by circles. If $\lambda > \lambda_{\mathrm{crit}}$, solitary wave solutions to \eqref{2.nkdv2} are exponentially localized in space. If $\lambda \leq \lambda_{\mathrm{crit}}$, the solutions are generalized solitary waves with non-decaying oscillations in the far field. Choices of $(k,\lambda)$ with even $k$ in the white region correspond to spatially localized solutions, and choices of $(k,\lambda)$ with even $k$ in the shaded region correspond to generalized solitary wave solutions. The 7KdV corresponds to $(k,\lambda_{\mathrm{crit}}) = (2,\tfrac{1}{4})$, whose value is indicated on each axis.}\label{F:lambdacrit}
\end{figure}

If $k$ is even, there exists some positive $\lambda_{\mathrm{crit}}$ such that if $\lambda > \lambda_{\mathrm{crit}}$, there are no imaginary solutions to \eqref{2.nsing2}, while if $\lambda \leq \lambda_{\mathrm{crit}}$, there are imaginary solutions for $\chi_x$. Consequently, the travelling wave solution to \eqref{2.nkdv2} has non-decaying far-field oscillations if $\lambda \leq \lambda_{\mathrm{crit}}$. Numerical investigation suggests that $\lambda_{\mathrm{crit}}$ increases as $k$ increases, and that $\lambda_{\mathrm{crit}} \rightarrow \tfrac{1}{2}$ as $k\rightarrow \infty$. Values of $\lambda_{\mathrm{crit}}$ are shown for a range of even $k$ in Figure \ref{F:lambdacrit}.

\begin{figure}
\begin{center}
\includegraphics{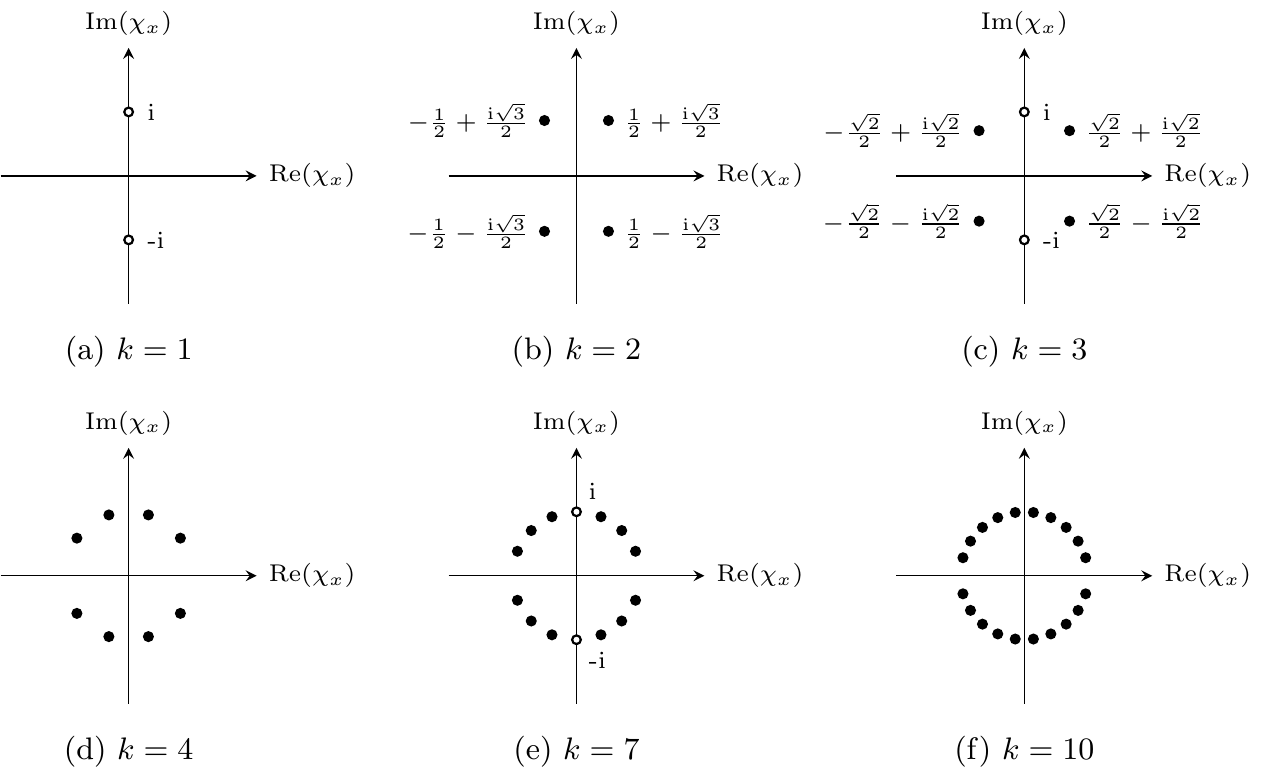}
\end{center}
\caption{Solutions to \eqref{2.nsing} for a range of $k$ values. Filled circles denote solutions with nonzero real part, and empty circles denote imaginary solutions.  The diagrams in (a) and (b) correspond to the singularly-perturbed fifth- and seventh-order KdV equations respectively. When $k$ is odd, there are solutions $\chi_x = \pm\mathrm{i}$, but there are no imaginary solutions if $k$ is even. Generalized solitary waves can only appear in systems with imaginary solutions, which correspond to odd values of $k$.}\label{F.nsols}
\end{figure}

In order to visualise the distribution of solutions for $\chi_x$ as $k$ varies, we consider the case of \eqref{2.nkdv2} for $\lambda = 1$. The corresponding singulant equation is given by
\begin{equation}
\sum_{r=0}^k \chi_x^{2r} = 0,\label{2.nsing}
\end{equation}
noting that $\chi_x \neq 0$. It is clear from direct factorization that when $k$ is odd, this polynomial has solutions given by $\chi_x = \pm \mathrm{i}$, while Figure \ref{F:lambdacrit} shows that this polynomial has no imaginary solutions when $k$ is even. Some examples of the solutions to this polynomial are seen in Figure \ref{F.nsols}.  In each case, solutions for $\chi_x$ that do not lie on the imaginary axis correspond to oscillations that grow or decay as $x - ct \rightarrow \pm\infty$, while the labelled solutions on the imaginary axis correspond to non-decaying oscillations in the far field.

Therefore we conclude that only when $k$ is odd can the higher-order KdV equation given in \eqref{2.nkdv2} for $\lambda = 1$ produce GSW. The corresponding solutions when $k$ is even must decay exponentially in the far field as $x - ct \rightarrow \pm\infty$. We do note from the pattern of solutions in Figure \ref{F.nsols} that for sufficiently large even $k$, the nearest $\chi_x$ that are nearest to the imaginary axis can be found arbitrarily close to $\chi_x = \pm \mathrm{i}$. As the real part of these solutions is very small, the spatial decay will be comparatively slow; however, the associated exponential behaviour will always decay away from $x = ct$.


\section{Finite-Difference KdV}\label{S:lKdV}
\subsection{Formulation and Series Expansion}

The previous analysis suggests that the exponential asymptotic analysis holds in a straightforward fashion, even as the order of the partial differential equation is taken to be arbitrarily large. In this case, the analysis is that of an infinite-order system, which is typically associated with a difference or delay equation, such as those considered in \cite{Joshi4, Joshi5, King4}. We therefore consider the behaviour of a \textit{lattice} KdV equation, generated by applying a finite difference scheme to the KdV equation.

Consider a finite-difference approximation of the KdV equation, with spatial grid distance given by $h$, and time step by $\tau$, where $h$ and $\tau$ are positive. Using central difference approximations for partial derivatives in time and space, we obtain a lattice version of the KdV equation (lKdV),
\begin{align}
\nonumber\frac{u(x+2h,t) - 2u(x+h,t) + 2u(x-h,t) - u(x-2h,t)}{2h^3}&\\
 + 6 u(x,t)\frac{u(x+h,t) - u(x-h,t)}{2h} +  &\frac{u(x,t+\tau) - u(x,t-\tau)}{2\tau}= 0.\label{1.fdkdv}
\end{align}
A similar analysis could be applied to other typical finite-difference approximations, such as using forward difference approximations for the time derivatives; the resultant analysis proceeds in a nearly-identical fashion.

We set $\tau = \sigma h$, and take the limit that $h \rightarrow 0$. Applying a Taylor approximation around $h = 0$ gives
\begin{align}
\nonumber\frac{1}{{h}^3}\sum_{r = 0}^{\infty} \frac{(2{h})^{2r+1}}{(2r+1)!} \pdiff{^{2r+1}u(x,t)}{x^{2r+1}}-\frac{2}{{h}^3}&\sum_{r = 0}^{\infty} \frac{{ h}^{2r+1}}{(2r+1)!}  \pdiff{^{2r+1}u(x,t)}{x^{2r+1}}\\
+ \frac{6u(x,t) }{h} \sum_{r = 0}^{\infty} \frac{{h}^{2r+1}}{(2r+1)!}& \pdiff{^{2r+1}u(x,t)}{x^{2r+1}}+ \frac{1}{\sigma h}\sum_{r = 0}^{\infty} \frac{(\sigma h)^{2r+1}}{(2r+1)!}  \pdiff{^{2r+1}u(x,t)}{t^{2r+1}} = 0.\label{1.fdkdv expanded}
\end{align}

We expand $u(x,t)$ as a power series in $h^2$, giving
\begin{equation}
u(x,t) \sim \sum_{j=0}^{\infty} h^{2j} u_j(x,t).\label{1.series}
\end{equation}

Applying this to \eqref{1.fdkdv expanded} gives
\begin{align}
\nonumber \sum_{r = 0}^{\infty}& \frac{2^{2r+1}h^{2r-2}}{(2r+1)!} \sum_{j=0}^{\infty}h^{2j} \pdiff{^{2r+1}u_j}{x^{2r+1}}-2\sum_{r = 0}^{\infty} \frac{h^{2r-2}}{(2r+1)!} \sum_{j=0}^{\infty}h^{2j} \pdiff{^{2r+1}u_j}{x^{2r+1}}\\
+& {6}\left[\sum_{j=0}^{\infty} h^{2j}u_j \right]\left[\sum_{r = 0}^{\infty} \frac{h^{2r}}{(2r+1)!} \sum_{j=0}^{\infty}h^{2j} \pdiff{^{2r+1}u_j}{x^{2r+1}}\right]+2\sum_{r = 0}^{\infty}\frac{{\sigma}^{2r}h^{2r}}{(2r+1)!}\sum_{j=0}^{\infty}h^{2j} \pdiff{^{2r+1}u_j}{t^{2r+1}}   = 0.\label{1.fdkdv series}
\end{align}
We see that, although we discretized an unperturbed KdV equation, the discretization process has introduced higher-order derivatives which disappear for $h=0$. Consequently the singular perturbation was created by the discretization process.

Matching this expression at $\mathcal{O}(1)$ as $h\rightarrow 0$ shows that $u_0$ is governed by the unperturbed KdV equation \eqref{0.kdv}. We therefore select as our leading-order solution the soliton solution of KdV, given in \eqref{2.u0}. We again denote the relevant singularities in the analytic continuation of $u_0(x,t)$ as $x_+$ and $x_-$, located at $x = ct \pm \mathrm{i} \pi/\sqrt{c}$.



To find subsequent terms in the series, we must match \eqref{1.fdkdv series} at higher orders of $h$. Matching at $\mathcal{O}(h^{2k})$ gives the equation
\begin{align}
\nonumber\sum_{r = 0}^{k} \frac{2^{2r+1}}{(2r+1)!} &\pdiff{^{2r+1}u_{k-r+1}}{x^{2r+1}} -2\sum_{r = 0}^{k} \frac{ 1}{(2r+1)!} \pdiff{^{2r+1}u_{k-r+1}}{x^{2r+1}}  = 0 
\\
 &- 6\sum_{j=0}^k\sum_{r=0}^{k-j} \frac{u_j }{(2r+1)!} \pdiff{^{2r+1}u_{k-r-j}}{x^{2r+1}}  +2\sum_{r = 0}^{k} \frac{{\sigma}^{2r}}{(2r+1)!}\pdiff{^{2r+1}u_{k-r}}{t^{2r+1}}= 0.\label{1.fdkdv recur}
\end{align}



\subsection{Late-Order Asymptotic Terms}

As before, this is a singularly-perturbed problem with singularities in the analytically continued leading-order behaviour, and that the series \eqref{1.series} will therefore diverge. We therefore again propose an ansatz for the late-order series terms of the form \eqref{2.ansatz}.


Substituting this ansatz into the recurrence relation \eqref{1.fdkdv recur} gives
\begin{align}
\nonumber \sum_{r = 0}^{k} \frac{2^{2r}-1}{(2r+1)!} \frac{F\chi_x^{2r+1} \Gamma(2k+3+\gamma)}{\chi^{2k+3+\gamma}}
\nonumber&- \sum_{r = 0}^{k} \frac{2^{2r}-1}{(2r)!} \frac{F_x\chi_x^{2r} \Gamma(2k+2+\gamma)}{\chi^{2k+2+\gamma}}
\\
+ \sum_{r = 0}^{k} &\frac{2^{2r}-1}{(2r)!} \frac{F\chi_x^{2r-1}\chi_{xx} \Gamma(2k+2+\gamma)}{\chi^{2k+2+\gamma}} + \ldots = 0,\label{1.late recur}
\end{align}
where the terms omitted are $\mathcal{O}(u_{k+1/2})$ in the limit as $k \rightarrow \infty$. 

Matching at $\mathcal{O}(u_{k + 3/2})$ as $k\rightarrow \infty$ gives
\begin{equation}
\sum_{r = 0}^{k}\frac{(2^{2r}-1)\chi_x^{2r + 1}}{(2r+1)!} = 0
\end{equation}
Recalling that the ansatz \eqref{2.ansatz} applies to late-order terms for which $k$ is large, we follow \cite{Joshi4,Joshi5,King4} and take the limit of the summation term as $k \rightarrow \infty$. This introduces only exponentially small error into the behaviour of $\chi$, which has no significant impact on the oscillatory tails in the solution. The resultant singulant equation is given by
\begin{equation}
[1-\cosh(\chi_x)]\sinh(\chi_x) = 0,\label{1.singulant0}
\end{equation}
or $\chi_x = \mathrm{i} \pi N$, where $N \in \mathbb{Z}$. Recalling that the singulant must be zero at the location of the leading-order singularities, denoted $x_s$, we solve \eqref{1.singulant0} to give
\begin{equation}
\chi = \mathrm{i} \pi N(x - x_s),\qquad N \in \mathbb{Z}.
\end{equation}
The dominant contributions to the late-order behaviour those associated with nonzero solutions of $\chi$ which have the smallest absolute value on the real axis, corresponding to $N = \pm 1$. This gives four possible late-order term contributions; however, a full analysis shows that only two of these will produce exponentially small contributions in the solution, associated with
\begin{equation} 
\chi_+ = \mathrm{i}  \pi (x- x_+),\qquad \chi_- = -\mathrm{i}\pi  (x - x_-).\label{1.singulant}
\end{equation}



Returning to \eqref{1.late recur} and matching at $\mathcal{O}(u_{k+1})$, noting that $\chi_{xx} = 0$, we find
\begin{equation}
\sum_{r = 0}^{k} \frac{(2^{2r}-1)F_x}{(2r)!} = 0.
\end{equation}
Again, we may take $k \rightarrow \infty$ while introducing only exponentially small error into the final contribution. We find that this gives $2F_x =0$. Consequently, $F$ is independent of $x$.

As in the analysis of 7KdV in Section \ref{S:LOT2}, we find that the ansatz can only be consistent with the leading-order solution near the singularity \eqref{2.singlocal} if $\gamma = 2$, so that the inner limit of the late-order ansatz is given by
\begin{equation}
u_j(x,t) \sim \frac{F(t)\Gamma(2j+2)}{[\mathrm{i} \pi (x-x_+)]^{2j+2}} \qquad \mathrm{as} \qquad x \rightarrow x_+ \quad \mathrm{and} \quad j \rightarrow \infty.\label{1.ansatzinner}
\end{equation}
A similar analysis applies to the late-order terms corresponding to the singularity at $x = x_-$.


We determine $F$ by matching the behaviour of the late-order terms with a local expansion in the neighbourhood of the singularity at $x_s$, in the same fashion as the 7KdV analysis. This inner analysis is performed in Appendix \ref{S.Inner}, and shows that the prefactors associated with both of $\chi_{\pm}$ are constant, both given by $F(x,t) = \Lambda \approx -2.68\times10^3$. 

The late order terms responsible for Stokes switching in the solution are therefore given by
\begin{equation}
u_j \sim \frac{\Lambda\Gamma(2j + 2)}{[\mathrm{i} \pi (x - x_+)]^{2j+2}} + \frac{ \Lambda\Gamma(2j + 2)}{[-\mathrm{i} \pi (x - x_-)]^{2j+2}},\qquad \mathrm{as} \qquad j \rightarrow \infty,\label{1.LOT}
\end{equation}
where $x_\pm = \pm \mathrm{i}\pi /\sqrt{c} + ct$.

\subsection{Exponential Asymptotics}

The exponential asymptotic analysis for the lattice KdV equation is almost identical to the analysis of 7KdV, however we demonstrate the analysis in order to illustrate the changes that occur due to the Taylor series terms in \eqref{1.fdkdv expanded}. 



The first step of the exponential asymptotic analysis is again to optimally truncate the divergent series. We apply the heuristic from \cite{Boyd1} and determine that $N \sim |\chi|/2h$ as $h \rightarrow 0$. We therefore set $N = |\chi|/2h + \omega$, where $0 \leq \omega < 1$ is chosen so that $N \in \mathbb{Z}$. We truncate the series \eqref{1.series} at this point to give
\begin{equation}
u(x,t) = \sum_{j = 0}^{N-1} h^{2j} u_j + R(x,t),
\end{equation}
where $R$ is the truncation error. 
%
We apply this expression to \eqref{1.fdkdv series}, and apply \eqref{1.fdkdv recur} to eliminate series terms. The largest remaining terms can be rearranged to give
\begin{align}
\nonumber 4&\sum_{r = 0}^{\infty} \frac{(2{h})^{2r-2}}{(2r+1)!} \pdiff{^{2r+1}R}{x^{2r+1}} -2\sum_{r = 0}^{\infty} \frac{{h}^{2r-2}}{(2r+1)!} \pdiff{^{2r+1}R}{x^{2r+1}}\\
 &+ 6\left[ R \pdiff{u_0}{x} + \ldots +\sum_{r = 0}^{\infty} \frac{{h}^{2r}}{(2r+1)!}\pdiff{^{2r+1}R}{x^{2r+1}}\right] +2\sum_{r = 0}^{\infty} \frac{\sigma^{2r}}{(2r+1)!}\pdiff{^{2r+1}R}{t^{2r+1}} = h^{2N}\partial_x^3 u_N + \ldots,\label{1.fdkdv truncated recur}
\end{align}
where again the omitted remainder terms are negligible in the limit $h \rightarrow 0$.

The right-hand side again only contributes to the remainder behaviour in the neighbourhood of the Stokes line.  We can therefore determine the behaviour away from the Stokes line by solving the homogeneous problem for $R$. Using a WKB (or Liouville-Green) ansatz, we again find the appropriate form for the remainder is given in the limit that $h \rightarrow 0$ as $R \sim \mathcal{S} \e^{-\chi/h}$.

Applying this expression to \eqref{1.fdkdv truncated recur} gives, after some simplification,
\begin{align}
\frac{2}{h^2}\sum_{r = 0}^{\infty} \frac{(2^{2r}-1)\chi_x^{2r}}{(2r)!}  \mathcal{S}_x \e^{-\chi/h}   \sim h^{2N}\partial_x^3 u_N, \qquad \mathrm{as} \qquad h \rightarrow 0.
\end{align}
Evaluating the summation terms for $\chi_x = \pm \mathrm{i} \pi$ and applying the late-order ansatz \eqref{2.ansatz} gives
\begin{align}
\frac{4}{{h}^2}\mathcal{S}_x  \e^{-\chi/h}  \sim -\frac{h^{2N}\chi_x^3  \Lambda \Gamma(2N + 5)}{\chi^{2N+5}},\label{1.Seq1}
\end{align}
where the omitted terms are smaller than those retained in the limit that $h \rightarrow 0$ and $N \rightarrow \infty$.

The remaining analysis is practically identical to Case B of the 7KdV analysis in Section \ref{S.CaseB}. We apply the late-order ansatz \eqref{2.ansatz} to \eqref{1.Seq1}, and and set the optimal truncation point to be $N = |\chi|/2h + \omega$. We transform into polar coordinates, so that $\chi = \rho \e^{\mathrm{i}\theta}$, and then consider the behaviour near $\theta = 0$, which corresponds to the Stokes line.

%

We again find that $\mathcal{S}_{\theta}$ is exponentially small except on the curve $\theta = 0$. Hence, this curve describes the Stokes line across which $\mathcal{S}$ varies rapidly, which has the same location as the Stokes curve in the fifth- and seventh-order KdV problems. The Stoke line behaviour is therefore identical to that shown in Figure \ref{F.7KdVstokes}, where $R^-$ and $R^+$ are given below, in \eqref{1.Rfkdv}.

We introduce inner coordinates $\theta = h^{1/2}\phi$, and solve the resultant expression. Converting back into outer coordinates gives
\begin{equation}
\mathcal{S} = -\frac{\mathrm{i}  \chi_x^2 \Lambda\sqrt{2\pi \rho}}{h^2 }\int_{-\infty}^{\sqrt{\rho}\theta/h} \e^{-s^2/2}\mathrm{d} s + \mathcal{S}_0,\label{1.Sint}
\end{equation}
where $\mathcal{S}_0$ is a constant of integration. Consequently, we see that the jump in $\mathcal{S}$ as the Stokes line is crossed from left to right is given by
\begin{equation}
\left[\mathcal{S}\right]_-^+ \sim -\frac{2\pi \mathrm{i} \chi_x^2 \Lambda}{h^2}.\label{1.sjump}
\end{equation}
The amplitude of the remainder is constant and exponentially small, and therefore does not decay or grow in the far field. As in Case B of the 7KdV analysis, we are free to choose $\mathcal{S}_0$ in \eqref{1.Sint}. Choosing this value so that $\mathcal{S} = 0$ on the Stokes line produces symmetric oscillatory behaviour about $x = ct$ gives 
%
%
%
\begin{equation}
R^- \sim -\frac{2\pi^3 \Lambda}{h^2}\e^{-\pi^2/h\sqrt{c}}\sin\left[\frac{\pi}{h}(x-ct)\right], \qquad R^+ \sim \frac{2\pi^3 \Lambda}{h^2}\e^{-\pi^2/h\sqrt{c}}\sin\left[\frac{\pi}{h}(x-ct)\right],\label{1.Rfkdv}
\end{equation}
where $R = R^-$ for $x < ct - \delta$, and $R = R^+$ for $x > ct + \delta$. These oscillations are symmetric about $x = ct$, as required. Consequently, Figure \ref{F.7KdVstokes} also shows the Stokes structure for the lattice KdV problem, where the remainder terms are non-decaying oscillatory functions given in \eqref{1.Rfkdv}. 

\begin{figure}
\centering
\includegraphics{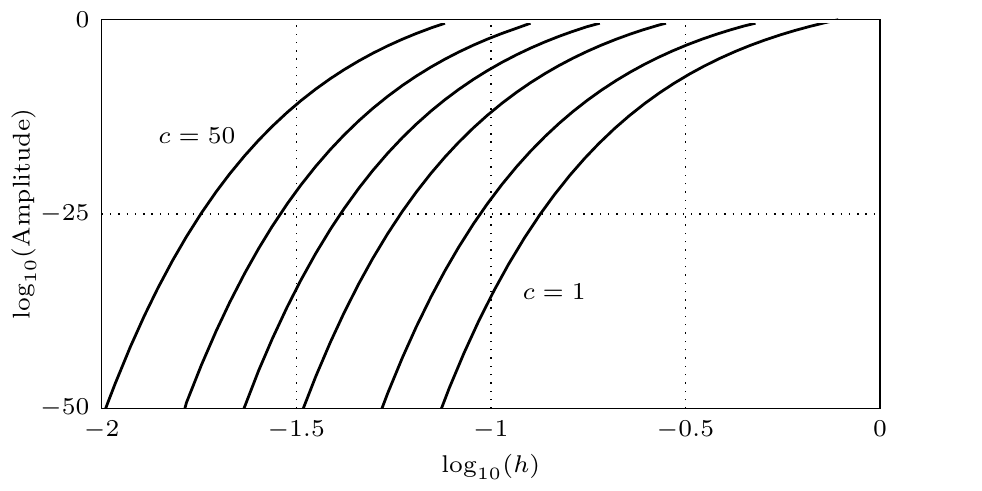}

\caption{Amplitude of far-field oscillations, calculated using \eqref{1.Rfkdv}, plotted logarithmically. From right to left, the curves correspond to $c = 1, 2, 5, 10, 20, 50$. The oscillations are exponentially small, but they do not decay in space or time.}\label{F:hamp}
\end{figure}

The amplitude of the far-field oscillations varies depending on the choice of $c$ and $h$, shown in Figure \ref{F:hamp}. The oscillations are clearly exponentially small in $h$, however they do not decay in the far field. 

\textcolor{black}{We again note that $\mathcal{S}_0$ could be chosen to enforce other conditions, such as requiring the system to be free of far-field oscillations either behind or ahead of the the point $x = ct$. }

\subsection{Lattice Equations with Parameters}\label{S:LattPar}

\textcolor{black}{It is straightforward to extend this methodology to higher-order variants of the KdV equation. In fact, the method is almost identical, as the resultant expression after applying a finite-difference discretization and then a Taylor series approximation is always a differential equation of infinite order, as in \eqref{1.fdkdv expanded}. Consequently, the only significant difference in approach when studying discretized versions of higher-order KdV equations is the algebraic complexity of the discretized expression.}

\textcolor{black}{Using this methodology, it is interesting to show that the process of discretization affects bifurcation parameters, such as those in \eqref{2.7KdV} or \eqref{0.kdv5}. Here, we consider discretization of the simplest KdV extension that contains a bifurcation parameter: the fifth-order KdV \eqref{0.kdv5}, which we will denote as 5KdV. We recall that Grimshaw \textit{et al.} \cite{Grimshaw6} showed that the value of $\kappa$ played a key role in the behaviour of the continuous problem \eqref{0.kdv5}, finding that the solution only contains GSW if $\kappa > 0$. Using a similar argument to Section \ref{S:Higher}, we will show that this is no longer true after discretization.}

\textcolor{black}{We apply a central difference discretization to the fifth-order KdV equation \eqref{0.kdv5}, using $\epsilon$ as the lattice interval in space and time. We then apply a nearly identical process to the analysis of lKdV in order to obtain a singulant equation, given by
\begin{equation}
\left[1-\cosh(\chi_x) + 8 \kappa \sinh(\chi_x/2)^4\right]\sinh(\chi_x) = 0.
\end{equation}
This equation has two sets of solutions, which we denote $\chi^{(1)}$ and $\chi^{(2)}$, that satisfy
\begin{equation}
\chi^{(1)}_x = \i \pi N, \qquad \chi^{(2)}_x = 4 \i \pi N \pm \log\left(\pm\sqrt{1- \frac{1\pm\sqrt{1-4\kappa}}{2\kappa}}\right),\label{3.sings}
\end{equation}
where $N \in \mathbb{Z}$, and the signs in $\chi_x^{(2)}$ may be chosen independently. }

\textcolor{black}{If $\kappa < \tfrac{1}{4}$, there are imaginary solutions of $\chi_x^{(2)}$. Hence, the GSW in the solution contain oscillatory terms associated with the singulant
\begin{equation}
\chi^{(1)} = \pm \i \pi (x-x_s),
\end{equation}
where $x_s$ is the singularity location, and we have restricted our attention to the dominant singulant contributions, given by $N = \pm 1$. These oscillations are clearly independent of $\kappa$. However, if $\kappa \geq \tfrac{1}{4}$, all values of $\chi_x^{(2)}$ are imaginary. In this case, the dominant singulant contributions are choices of $\chi^{(2)}$ with $N=0$ and the first sign choice in \eqref{3.sings} taken as positive. This gives, after some simplification
\begin{equation}
\chi^{(2)} = \pm  \frac{1}{2}\log\left({1- \frac{1\pm\sqrt{1-4\kappa}}{2\kappa}}\right) (x-x_s).
\end{equation}
In this case, the behaviour of the far-field oscillations does depend on $\kappa$. }
\textcolor{black}{
Hence we see that, unlike solutions to the corresponding continuous fifth-order KdV equation, the asymptotic solution to the discretized version of 5KdV contains exponentially small oscillations in the far field for any choice of $\kappa$. There is, however, a critical value of this parameter, $\kappa_{\mathrm{crit}} = \tfrac{1}{4}$, at which the behaviour of the far-field oscillations of the GSW changes. If $\kappa < \kappa_{\mathrm{crit}}$, the singulant (and hence the oscillation frequency) does not depend on $\kappa$; however, if $\kappa \geq \kappa_{\mathrm{crit}}$, the dominant oscillatory behaviour is dependent on the value of $\kappa$. Consequently, crossing the critical parameter value does not switch off the oscillations entirely, as in the continuous problem, but rather changes their form.}

We conclude that the process of discretization can cause significant changes in the behaviour of bifurcation parameters in systems such as 5KdV. An equivalent process may be applied to discretized versions of higher-order KdV extensions from the hierarchy \eqref{2.nkdv2}, such as 7KdV. For these problems, the asymptotic analysis is nearly identical to the analysis of 5KdV, and the critical parameter value is associated with a change in oscillation regimes, rather than changing between classical and generalized solitary wave solutions. As the asymptotic analysis of later equations in the KdV hierarchy from \eqref{2.nkdv2} is nearly identical to the analysis in this section, the details are not presented here.

\section{Discussion and Conclusions}\label{S:Discuss}

The purpose of this study was to identify instances of GSW in a lattice equation. We considered a finite difference discretization scheme to the unperturbed KdV equation \eqref{0.kdv} and utilized Taylor expansions in order to represent the lattice equation in terms of an infinite sum of derivative terms. We then applied methods developed in \cite{Joshi4, Joshi5, King4} for difference equations in order to study exponentially small contributions to the solution behaviour. We found that the solution contains Stokes lines across which non-decaying exponentially small oscillations are switched. This demonstrates that lattice equations can produce GSW solutions in the asymptotic limit corresponding to the discretization parameter, even if the system being discretized is not singularly perturbed. We also showed that the same process can be applied to higher order lattice equations. By studying the lattice version of the singularly-perturbed fifth-order KdV, we demonstrated that the discretization can alter the both the critical value and effects of bifurcation parameters in the system.

We first motivated this result by considering exponentially small terms in the seventh-order KdV equation \ref{2.7KdV}, and subsequently a higher-order KdV hierarchy \eqref{2.nkdv2}. This extends the work of \cite{Grimshaw2,Grimshaw1,Trinh5}, who considered the singularly perturbed fifth-order KdV using exponential asymptotic methods. We first identified that the exponentially small oscillations in 7KdV have constant, non-decaying amplitude in the far field for $\lambda \leq \tfrac{1}{4}$. If instead $\lambda > \tfrac{1}{4}$, then the oscillations can only decay in the far field, meaning that the wave is localized, and the solution is a classical solitary wave. This is consistent with the result obtained by \cite{Pomeau1}. We then found that solutions of the KdV hierarchy also critical values of $\lambda$ at which the solution changed between GSW and classical solitary waves. For even $k$, corresponding to singularly-perturbed differential equations with order $4\mathbb{Z}+1$, we found that travelling wave solutions for $\lambda > 0$ were GSW, while solutions for $\lambda < 0$ were localized solitary waves. This result is consistent with the work of \cite{Grimshaw6} for the fifth-order KdV in this hierarchy.

\textcolor{black}{An important consequence of this work is that discretized systems can potentially produce exponentially small oscillations in the continuum limit as a consequence of the discretization process. We note that the Taylor approximation for $f(x + a h)$ about $h = 0$ is written as 
\begin{equation*}
\sum_{r=0}^{\infty}\frac{(ah)^r}{r!} \frac{\partial^r f(x)}{\partial x^r}.
\end{equation*} 
Finite difference approximations are composed of a sum of terms with advances and delays, and hence applying a Taylor series approximation to a finite-difference system must produce a differential equation of infinite order that is singularly perturbed in the lattice continuum limit $h \rightarrow 0$, such as \eqref{1.fdkdv expanded}. This analysis demonstrates the far-reaching consequence that exponentially small solution effects can appear in the continuum limit due to the discretization process, which produces a singularly-perturbed system even in cases where the equation does not have a natural small parameter or obvious singular perturbation. Similar effects also be seen in asymptotic solutions of difference equations, as in \cite{Joshi4, Joshi5}. }

While the amplitude of oscillations introduced by discretization is typically extremely small (see Figure \ref{F:hamp}), these could cause difficulties in the study of systems such as the 7KdV, which can have oscillations of comparably small size (see Figure \ref{F.7KdVR}). It is apparent from the oscillation amplitude in Figure \ref{F:hamp} that this difficulty can be averted by selecting appropriately small values of $h$. This value should be chosen in order to ensure that any oscillations caused by the discretization of the system are much smaller than any other feature of interest in the solution, such as non-decaying oscillatory behaviour that is present in the asymptotic solution to the original, non-discretized problem.

There are a number of extensions to the problems considered here that warrant closer examination. It would be of interest to understand how GSW in these systems interact. The problems considered here concentrated on the one-soliton solution \ref{2.u0}, however it would be of interest to see if more complicated behaviours and interaction effects arise as a consequence of wave interactions in $n$-soliton solutions. In particular, such solutions would allow us to identify bound states, as in \cite{Calvo1, Grimshaw4}, in which the oscillations that are switched on by one solitary wave are switched off by another. This means that, even though the solutions are GSW, the waves are not necessarily present in the far field, but rather restricted to the region between to travelling wave cores.

It would also be of great interest to study the behaviour of solutions in lattice equations that are not travelling waves, but rather the evolution of some prescribed initial condition. In the study of partial differential equations using exponential asymptotic methods \cite{Chapman4, Howls1}, complicated Stokes switching effects such as higher-order and second-generation Stokes switching. These effects have never been computed in lattice equations, and it would be of significant interest to determine whether such effects are present in lattice equation solutions which are allowed to evolve from general initial data.

\textcolor{black}{One problem not considered in this paper is the interaction of small parameters. In our analysis of the discretization of 5KdV found in Section \eqref{S:LattPar}, we selected the small parameter $\epsilon$, contained in the original problem, as our lattice parameter. However, if we chose the lattice distance to be an independent small quantity denoted by $h$, we could study the $h \ll \epsilon$ and $\epsilon \ll h$ cases separately. If $\epsilon \ll h$, the leading-order oscillations in the asymptotic solution must be identical to those found by the lKdV analysis from Section \ref{S:lKdV}. If $h \ll \epsilon$, the asymptotic lattice behaviour would be obtained by instead first discretizing the full fifth-order system. It would be particularly interesting to study the distinguished limit $h = \mathcal{O}(\epsilon)$ and show how it connects the two regimes.  A study of these individual cases would provide insight into how best to formulate singularly-perturbed systems using finite-difference methods in order to ensure that the behaviour caused by the singular perturbation is captured correctly in the discretized equation.}

Furthermore, there are a number of other problems that could be studied using the methods described in the present study. It would be of particular interest to study oscillations present in singularly perturbed variants of higher-order integrable variants of the KdV equation, such as the seventh-order Lax equation \cite{Lax1}, or members of the classical KdV hierarchy described in \cite{Newell1}. The unperturbed versions of these equations possess solitary wave solutions; it would be interesting to study the effect of discretization on these systems, and whether applying a Taylor series expansion to the discretized equations would produce GSW solutions.

It would also be worthwhile to apply exponential asymptotic methods to integrable discretizations of the KdV equations, such as the lattice KdV derived in \cite{Nijhoff1} and studied at length in \cite{Nijhoff2}. There exist many other integrable discretizations of KdV variants, catalogued in \cite{Hietarinta1}. As these discrete systems are integrable, they permit classical solitary wave (or plane wave) solutions. This suggests that the process of discretization process does not necessarily produce GSW solutions in the manner seen for the non-integrable hierarchy \eqref{2.nkdv2}, and it would be of interest to understand how this occurs. 
%
%
%
%

\section{Acknowledgements}

NJ was supported by Australian Laureate Fellowship grant no. FL120100094 from the Australian Research Council. CJL was supported by Macquarie University New Staff grant no. 63934274.

\bibliography{sydrefs2.bib}
\bibliographystyle{amsplain}

\appendix
\section{Inner Analysis for the Seventh-Order KdV}\label{S.Inner0}

In order to determine the behaviour of $F(x,t)$, we need to match the late-order ansatz \eqref{2.ansatz} to the inner problem near the singularity at $x = x_{\pm}$. 

We define an inner spatial variable $\epsilon \eta = x - x_{\pm}$, and a new independent coordinate $ v(\eta,t) = \epsilon^2 u(x,t)$. The governing equation \eqref{2.7KdV} becomes
\begin{align}
\frac{\lambda}{\epsilon^3}\pdiff{^7 v(\eta,t)}{\eta^{7}}+\frac{1}{\epsilon^3}\pdiff{^5 v(\eta,t)}{\eta^{5}}+\frac{1}{\epsilon^3}\pdiff{^3 v(\eta,t)}{\eta^{3}} + \frac{6v(\eta,t)}{\epsilon^3}\pdiff{v(\eta,t)}{x} + \frac{1}{\epsilon^2}\pdiff{v(\eta,t)}{t}=0\label{2.kdvinner}
\end{align}
In order to match the outer behaviour with the inner region, we require only the leading order governing equation for the inner region behaviour in the limit that $\epsilon \rightarrow 0$, which contains terms that are $\mathcal{O}(1/\epsilon^3)$. The inner expansion therefore satisfies
\begin{align}
\lambda \pdiff{^7 v}{\eta^{7}}+\pdiff{^5 v}{\eta^{5}}+\pdiff{^3 v}{\eta^{3}} + {6v}\pdiff{v}{x}= 0.\label{2.kdvinner leading}
\end{align}
We note that time derivatives do not contribute at this order, and therefore $v_j$ is constant in time. We now apply an asymptotic expansion in the inner coordinate, in the limit that $\eta \rightarrow \infty$, given by
\begin{equation}
v(\eta,t) = \sum_{j=0}^{\infty}\frac{v_j}{\eta^{2j+2}}.\label{2.innerseries}
\end{equation}
We set $v_0$ = -2 in order for this expression be consistent with the local singularity behaviour described in \eqref{2.singlocal}. We match this expression at successive orders of $\eta$ in order to obtain a recurrence relation for $v_j$. By matching at $\mathcal{O}(1/\eta^{2k+5})$ and simplifying the expression, we obtain the recurrence relation
\begin{equation}
\left[\frac{(2k+4)!}{(2k+1)!}+12k v_0\right]v_k = \frac{\lambda(2k+4)!v_{k-2}}{(2k-3)!}+\frac{(2k+4)!v_{k-1}}{(2k-1)!}+ 6\sum_{r=1}^{k-1}(2k-2r+2)v_r v_{k-r},\label{2.innerrecur}
\end{equation}
where we use the convention that $v_j=0$ for $j < 0$. This expression can be rearranged to obtain a recurrence relation for $v_{k}$ in terms of $v_j$ for $j < k$, beginning with $v_0 = -2$. By repeatedly applying this recurrence relation, values of $v_j$ may be obtained for any choice of $j$.

Matching the inner expansion \eqref{2.innerseries} with the outer ansatz \eqref{2.ansatzinner} requires some care, as we must account for both of the contributions that dominate behaviour in the neighbourhood of $x = x_+$. The full inner limit of the outer expansion \eqref{2.ansatzinner} is therefore 
\begin{equation}
u_j(x,t) \sim \frac{F_1(t)\Gamma(2j+2)}{[\alpha (x-x_+)]^{2j+2}}+\frac{F_2(t)\Gamma(2j+2)}{[-\overline{\alpha} (x-x_+)]^{2j+2}},\label{2.limit inner}
\end{equation}
as $j \rightarrow \infty$ and $x \rightarrow x_+$, where $\alpha = \tfrac{1}{2} + \tfrac{\mathrm{i}\sqrt{3}}{2}$. Matching between the outer limit of the inner expansion \eqref{2.innerseries} and the inner limit of the outer expansion \eqref{2.limit inner} is complicated by the fact that there are two contributions of equal size.

We follow the strategy for finding the prefactor with multiple contributions used in \cite{Alfimov2, Joshi4, Joshi5} and adapt it to consider the sum of three consecutive terms. Consequently, we find that
\begin{equation}
3F_1(t) = \lim_{j\rightarrow \infty}\left[\frac{v_j \alpha^{2j+2}}{\Gamma(2j+2)}  +\frac{v_{j+1} \alpha^{2j+4}}{\Gamma(2j+4)} + \frac{v_{j+2} \alpha^{2j+6}}{\Gamma(2j+6)}\right],\label{A.F1}
\end{equation}
where the fact that $1 + (\alpha/\overline{\alpha})^2 + (\alpha/\overline{\alpha})^4 = 0$ ensures that all terms containing $F_2$ are eliminated from the expression. If we instead multiply throughout by $-\overline{\alpha}$, we similarly find that
\begin{equation}
3F_2(t) = \lim_{j\rightarrow \infty}\left[\frac{v_j \overline{\alpha}^{2j+2}}{\Gamma(2j+2)}  +\frac{v_{j+1} \overline{\alpha}^{2j+4}}{\Gamma(2j+4)} + \frac{v_{j+2} \overline{\alpha}^{2j+6}}{\Gamma(2j+6)}\right].\label{A.F2}
\end{equation}

\begin{figure}
\centering
\includegraphics{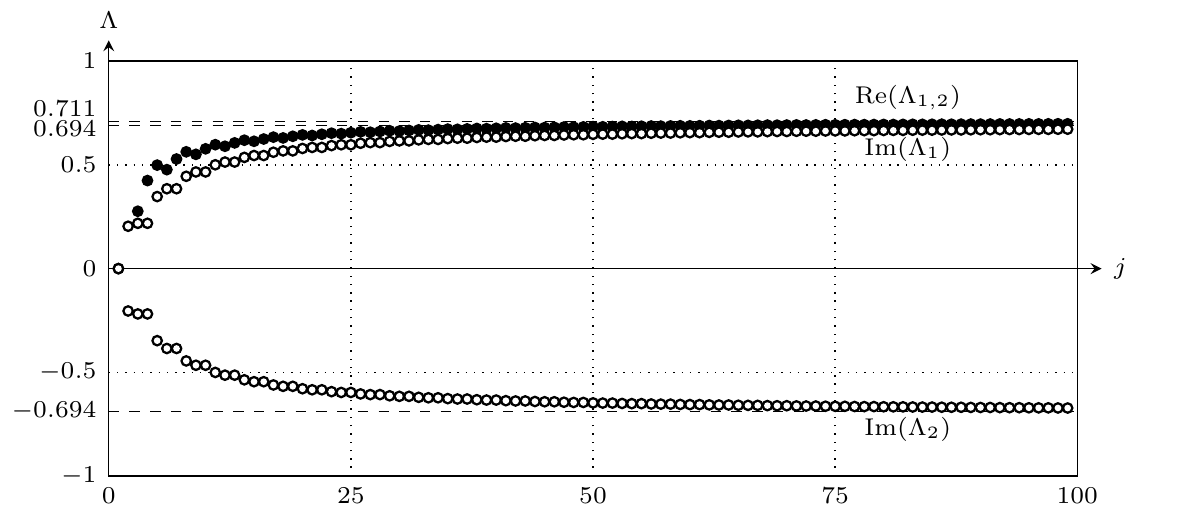}
\caption{Approximation for $\Lambda_1$ and $\Lambda_2$ obtained by computing the expression in \eqref{A.F1} and \eqref{A.F2} for increasing values of $j$. The real part and imaginary parts of $\Lambda_i$ for $i = 1,2$ are represented by black and white circles respectively. }\label{F:Lambda0}
\end{figure}

As $v_j$ does not vary in time, $F_1$  and $F_2$ clearly take constant values. We will denote this value as $F = \Lambda_1$ and $\Lambda_2$ respectively. By evaluating $v_j$ for sufficiently large values of $j$ using \eqref{2.innerrecur}, we can approximate $\Lambda_i$ for $i = 1,2$ using \eqref{A.F1} and \eqref{A.F2}. This process is illustrated in Figure \ref{F:Lambda0} for $\lambda = 1$.

We see that as $v_j$ increases, $\Lambda$ tends to a constant value. This process was continued to $j = 600$, establishing that if $\lambda = 1$, we have $\Lambda_1\approx 0.711 + 0.694\mathrm{i}$ and $\Lambda_2 = \overline{\Lambda_1}$. A similar analysis in the neighbourhood of $x = x_-$ shows that $\Lambda_3 = \overline{\Lambda_1}$ and $\Lambda_4 = \overline{\Lambda_2}$; this is a  consequence of the fact that $u_j$ and $v_j$ are always real, and therefore must be the sum of complex conjugates. We will subsequently denote $\Lambda_1 = \Lambda$, giving $\Lambda_3 = \Lambda$ and $\Lambda_2 = \Lambda_4 = \overline{\Lambda}$.

%
 
\begin{figure}
\centering
%
%
%
%
%
%
%
%
\includegraphics{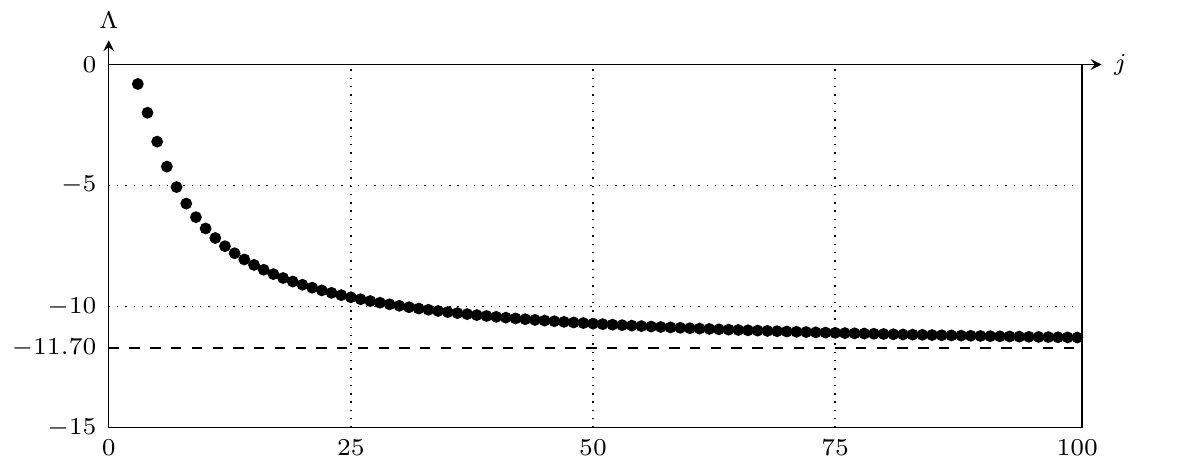}

\caption{Approximation for $\Lambda_+$ obtained by computing the expression in \eqref{2.F3}  for increasing values of $j$.}\label{F:Lambda02}
\end{figure}

In the $\lambda \leq \tfrac{1}{4}$ case, matching the inner expansion \eqref{2.innerseries} with the outer ansatz \eqref{1.ansatzinner} simply requires that
\begin{equation}
\Lambda_+ = \lim_{j\rightarrow \infty} \frac{v_j (\mathrm{i} \beta)^{2j+2}}{\Gamma(2j+2)},\label{2.F3}
\end{equation}
where $\Lambda_+$ is the constant associated with the singularity at $x = x_+$. This is always a real-valued expression, meaning that unlike the $\lambda > \tfrac{1}{4}$ case, $\Lambda$ will only take real values. An example calculation for $\lambda = \tfrac{1}{8}$ using the recurrence relation \eqref{2.innerrecur} is illustrated in Figure \ref{F:Lambda02}. The value of $\Lambda_+$ was approximated using \eqref{F:Lambda0} with $j = 600$, giving $\Lambda \approx -11.70$. A similar analysis shows that the corresponding constant associated with the singularity at $x = x_-$, denoted $\Lambda_-$ is identical to $\Lambda_+$. We therefore denote both using the constant $\Lambda$.

\section{Inner Analysis for the Lattice KdV}\label{S.Inner}

In order to determine the behaviour of $F(x,t)$, we need to match the late-order ansatz \eqref{2.ansatz} to the inner problem near the singularity at $x = x_{\pm}$. The process is similar to Appendix \ref{S.Inner0}, although the recurrence relation is more complicated, as the number of terms depends on the recursion depth. 

We define an inner spatial variable $h \eta = x - x_{\pm}$, and a new independent coordinate $ v(\eta,t) = h^2 u(x,t)$. The governing equation \eqref{1.fdkdv expanded} becomes
\begin{align}
\nonumber \frac{1}{h^3}\sum_{r = 0}^{\infty} \frac{2^{2r+1}}{(2r+1)!}& \pdiff{^{2r+1} v(\eta,t)}{\eta^{2r+1}} -\frac{2}{h^3}\sum_{r = 0}^{\infty} \frac{1}{(2r+1)!} \pdiff{^{2r+1} v(\eta,t)}{\eta^{2r+1}} \\
+& \frac{6 v(\eta,t)}{h^3 }  \sum_{r = 0}^{\infty} \frac{1}{(2r+1)!}\pdiff{^{2r+1} v(\eta,t)}{\eta^{2r+1}}+  \frac{1}{h^2}\sum_{r = 0}^{\infty} \frac{(\sigma h)^{2r}}{(2r+1)!} \pdiff{^{2r+1} v(\eta,t)}{t^{2r+1}} = 0.\label{1.fdkdv inner}
\end{align}
The matching requires only the leading-order terms the limit that $h \rightarrow 0$. At $\mathcal{O}(1/h^3)$, we have
\begin{align}
\sum_{r = 0}^{\infty} &\frac{2^{2r+1}}{(2r+1)!}  \pdiff{^{2r+1} v}{\eta^{2r+1}}-2\sum_{r = 0}^{\infty} \frac{1}{(2r+1)!}  \pdiff{^{2r+1} v}{\eta^{2r+1}} +6v \sum_{r = 0}^{\infty} \frac{1}{(2r+1)!} \pdiff{^{2r+1} v}{\eta^{2r+1}}= 0.\label{1.fdkdv inner leading}
\end{align}
The time derivatives again do not contribute at this order, and therefore $v_j$ is a constant. We now apply an asymptotic expansion in the inner coordinate, in the limit that $\eta \rightarrow \infty$, given by \eqref{2.innerseries}.
We again set $v_0$ = -2 in order for this expression be consistent with the local singularity behaviour described in \eqref{2.singlocal} and match $\eta\rightarrow\infty$ in order to obtain a recurrence relation for $v_j$. At $\mathcal{O}(1/\eta^{2k+5})$, we obtain the recurrence relation
\begin{equation}
\sum_{r=1}^{k+1} {{2k+4}\choose{2r+1} }(2^{2r}-1)v_{k-r+1}+3\sum_{l=0}^{k}\sum_{r=0}^{l} {{2k+4}\choose{2r+1}} v_{k-l}v_{l-r} =0.\label{1.innerrecur}
\end{equation}
This expression can be rearranged to obtain a recurrence relation for $v_{k}$ in terms of $v_j$ for $j < k$, beginning with $v_0 = -2$. By repeatedly applying this recurrence relation, values of $v_j$ may be obtained for any choice of $j$.

\begin{figure}
\centering
%
%
%
%
%
%
%
%
%
\includegraphics{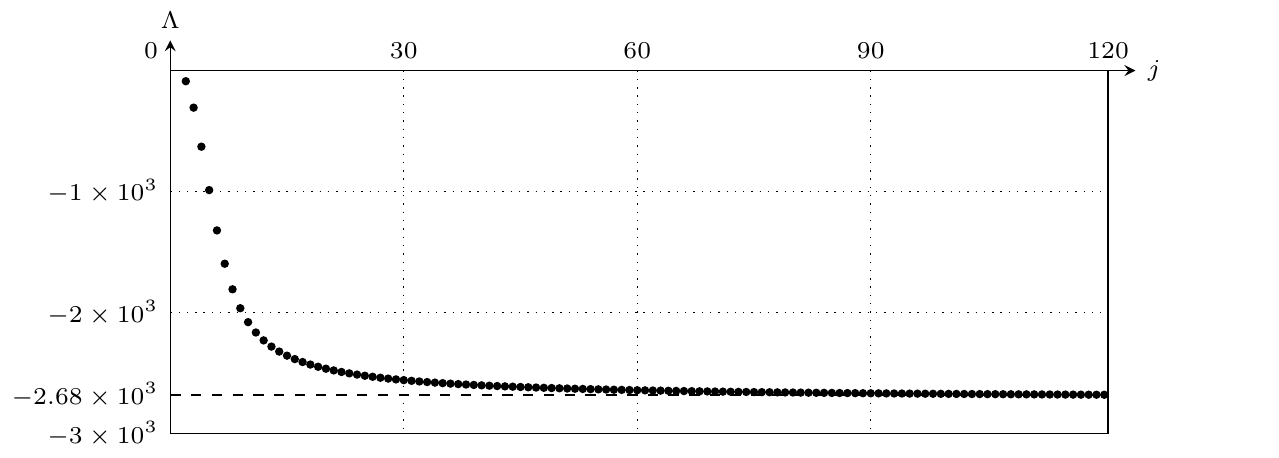}

\caption{Approximation for $\Lambda$ obtained by computing the expression in \eqref{1.limit inner} for increasing values of $j$.}\label{F:Lambda1}
\end{figure}

Matching the inner expansion \eqref{2.innerseries} with the outer ansatz \eqref{1.ansatzinner} requires that
\begin{equation}
F(t) = \lim_{j\rightarrow \infty} \frac{v_j (\mathrm{i} \pi)^{2j+2}}{\Gamma(2j+2)}.\label{1.limit inner}
\end{equation}
As $v_j$ is constant, $F(t)$ must also be a constant, which we again denote as $\Lambda$. By evaluating $v_j$ for sufficiently large values of $j$ using \eqref{1.innerrecur}, we can approximate $\Lambda$ using \eqref{1.limit inner}. This process is illustrated in Figure \ref{F:Lambda1}. As $v_j$ increases, the approximation for $\Lambda$ tends to a constant value. This process was continued to $j = 200$, establishing that $\Lambda\approx -2.68\times10^3$.

\end{document}